# A 5.3 GHz $Al_{0.76}Sc_{0.24}N$ Two-Dimensional Resonant Rods Resonator with a $k_t^2$ of 23.9%

Xuanyi Zhao, Onurcan Kaya, Michele Pirro, Meruyert Assylbekova
Luca Colombo, Pietro Simeoni, and Cristian Cassella
*Northeastern University*, Boston (MA), USA

*Abstract*—This work reports on the measured performance of an Aluminum Scandium Nitride (AlScN) Two-Dimensional Resonant Rods resonator (2DRR), fabricated by using a Sc-doping concentration of *24%*, characterized by a low off-resonance impedance (~25 Ω) and exhibiting a record electromechanical coupling coefficient ($k_t^2$) of *23.9%* for AlScN resonators. In order to achieve such performance, we identified and relied on optimized deposition and etching processes for highly-doped AlScN films, aiming at achieving high crystalline quality, low density of abnormally oriented grains in the 2DRR's active region and sharp lateral sidewalls. Also, the 2DRR's unit-cell has been acoustically engineered to maximize the piezo-generated mechanical energy within each rod and to ensure a low transduction of spurious modes around resonance. Due to its unprecedented $k_t^2$, the reported 2DRR opens exciting scenarios towards the development of next generation monolithic integrated radio-frequency (RF) filtering components. In fact, we show that 5$^{th}$-order 2DRR-based ladder filters with fractional bandwidths (BW) of *~11%*, insertion-loss (I.L) values of *~2.5 dB* and with *>30 dB* out-of-band rejections can now be envisioned, paving an unprecedented path towards the development of ultra-wide band (UWB) filters for next-generation Super-High-Frequency (SHF) radio front-ends.

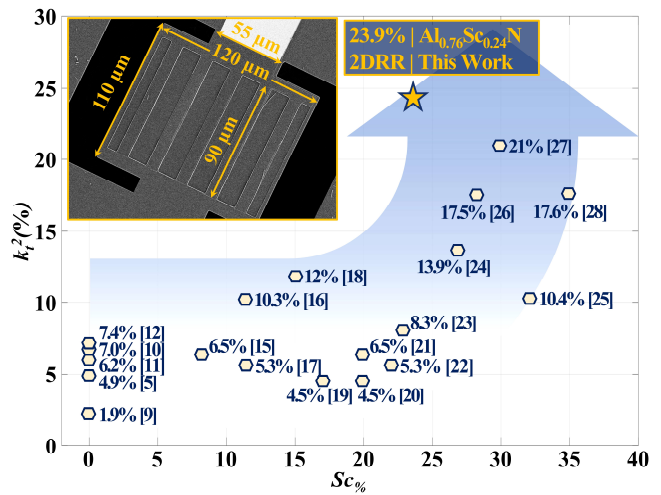

Figure 1: A comparison of the measured $k_t^2$ of the AlN and AlScN μARs reported to date *vs.* their adopted $Sc_\%$ value. The plot also includes the $k_t^2$ we measured for the AlScN 2DRR reported in this work (see the yellow star marker). A Scanned Electron Microscope (SEM) picture of the same 2DRR is shown in the in-set.

## I. INTRODUCTION

In the last two decades, microacoustic resonators (μARs) have played a key role in integrated *1G*-to-*4G* radios, providing the technological means to achieve compact radio-frequency (RF) filters with low loss and moderate fractional bandwidths (BW<*4%*). More specifically, Aluminum Nitride (AlN) based filters have populated the front-end of most commercial mobile transceivers due to the good dielectric, piezoelectric and thermal properties exhibited by AlN thin-films and because their fabrication process is compatible with the one used for any Complementary Metal Oxide Semiconductor (CMOS) integrated circuits (*IC*s) [1][2]. Nevertheless, the rapid growth of *5G* and the abrupt technological leap expected with the development of sixth-generation (*6G*) communication systems are expected to severely complicate the design of future radio front-ends by demanding Super-High-Frequency (SHF) filtering components with much larger fractional bandwidths than achievable today. As the bandwidth of any acoustic filter is directly proportional to the electromechanical coupling coefficient ($k_t^2$) exhibited by its forming resonators, a large effort has been recently made to identify new μAR-designs or alternative piezoelectric materials granting larger $k_t^2$ than what attained by the existing counterparts, all relying on the transduction of *Lamb modes* [3] in piezoelectric plates. Even more, since the acoustic filter topologies granting the highest BW leverage a set of electrically coupled μARs with different resonance frequencies, a large attention has also been paid to identify new acoustic technologies allowing to monolithically integrate μARs with different lithographically defined resonance frequencies ($f_{res}$), without increasing fabrication complexity and costs. In this regard, new AlN μARs have been recently reported, still relying on the transduction of acoustic waves in plates but achieving boosted $k_t^2$ values by leveraging: *i*) a segmented electrode excitation [4], *ii*) the excitation of dispersive $S_1$ modes [5][6], *iii*) the transduction of two-dimensional modes [7] around the dilatation frequency or *iv*) the spatial sampling of *piston*-like displacement modal distributions through an engineered dispersion of simultaneously transduced Lamb modes [8]. Nevertheless, while these new μARs enable a lithographic frequency tunability and a boosted electromechanical coupling coefficient, they can only ensure $k_t^2$ values approaching but not exceeding what attained by state-of-the-art AlN devices, such as Film-Bulk-Acoustic-Resonators (FBARs [10]). Differently, a larger $k_t^2$ than what attained by AlN FBARs has been enabled through the recent invention of Two-Dimensional-Resonant-Rods resonators (2DRRs [12][44]). 2DRRs leverage the exotic dispersive features of acoustic metamaterials, built out of thin-film corrugated piezoelectric layers, to simultaneously achieve a significant lithographic frequency tunability (>*10%*) and a $k_t^2$ exceeding what possible when relying on un-corrugated piezoelectric plates, like the ones used to build FBARs. Nevertheless, the maximum theoretical $k_t^2$ value of AlN 2DRRs (~*9%*) remains severely limited by the AlN piezoelectric coefficients, making AlN 2DRRs not usable to make ultra-wideband (UWB) filters for next-generation SHF front-ends. For this reason, in this work we designed, fabricated and tested a 2DRR relying on Aluminum Scandium Nitride (AlScN [13]), a piezoelectric material attained by doping AlN with scandium dopants and characterized by piezoelectric coefficients growing proportionally to the scandium-doping concentration ($Sc_\%$). More specifically, the 2DRR reported in this work relies on a *24% $Sc_\%$* value, it has a resonance frequency ($f_{res}$) of *~5.3 GHz* and it shows the highest $k_t^2$ (*23.9%*) ever demonstrated in AlN or AlScN µARs [5-28]. A comparison of the $k_t^2$ achieved by the 2DRR reported in this work with what attained by the previously reported highest-$k_t^2$ AlN and AlScN counterparts is provided in Figure 1. It is worth emphasizing that the 2DRR reported here achieves a $k_t^2$ value that is even higher than what attained by AlScN FBARs (*21%*) exploiting higher $Sc_\%$ values [27]. In order to achieve such unprecedented electromechanical performance in the reported 2DRR, we developed optimized deposition and etching recipes for highly doped AlScN films, aiming at achieving the best possible film crystallinity, the lowest density of AlScN abnormally oriented grains (AOGs) in the 2DRR's active region and a low residual stress, along with sharp sidewalls after the AlScN etch. Even more, we engineered the 2DRR's unit-cell to *trap* the piezoelectrically generated acoustic energy within each rod, and to suppress any undesired spurious modes affecting the transduction efficiency and the spectral purity of the 2DRR's frequency response. It is important to point out that the $k_t^2$ exhibited by the 2DRR reported in this work is also one of the highest ones ever reported for any SHF µARs, including the ones leveraging piezoelectric materials, like LiNbO$_3$, that are not manufacturable through CMOS-compatible fabrication processes (Table I).

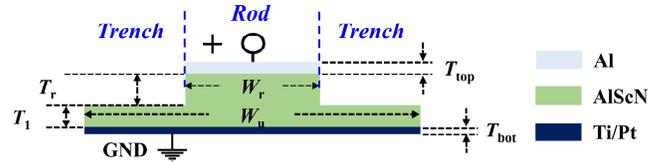

Figure 2: Schematic view of a unit-cell of the reported 2DRR. For the reported 2DRR, each unit-cell uses a rod width ($W_r$) of *9 µm*, a unit-cell width ($W_u$) of *24 µm*, a $T_r$ of *350 nm* and a $T_1$ of *150 nm*. Thicknesses for the (Ti/Pt) bottom metal layer and for the top (Al) metal strips are *20/50 nm* and *100 nm*, respectively. With regard to the reported 2DRR's excitation strategy, the top metal strips are all connected to the same voltage polarity, whereas the bottom plate is grounded.

In the following, we will first present the design and main operational features of the reported 2DRR by means of Finite Element Methods (FEM). We will numerically show how the proper engineering of the 2DRR's unit-cell permits to generate *stop-bands*, inhibiting the acoustic propagation out of each piezo-active rod. This provides key means to really leverage the high electromechanical performance of 2DRRs, as well as to improve the spectral purity of their frequency response by suppressing the propagation of Lamb waves moving across the 2DRR's corrugated film. Then, we will describe the complete fabrication process we followed to build the reported 2DRR. More specifically, we will discuss the processing conditions we identified and relied on to deposit and etch the Al$_{0.76}$Sc$_{0.24}$N film used by the reported 2DRR. Also, we will show the results of a complete set of material characterization experiments, demonstrating the high-quality of the sputtered and etched AlScN film used to build the 2DRR presented here. Later, we will showcase the 2DRR's electrical performance, extracted by a direct measurement through conventional RF characterization tools. Finally, given the 2DRR's exceptional electromechanical performance demonstrated in this work, we will discuss the new unveiled potential for building future SHF acoustic filters, with unprecedented ultra-wide BWs and low insertion-loss values.

## II. 2DRRs - PRINCIPLE OF OPERATION

A 2DRR consists of a group of identical unit-cells, each one formed by one piezo-active *rod* symmetrically placed between two identical *trenches* (see Figure 2). Each trench relies on a thin piezoelectric layer (with thickness $T_1$), deposited onto a full grounded metal plate. The rods, instead, use a thicker piezoelectric layer with thickness $T_r$. Also, the portion of the piezoelectric layer used for the rods is fully covered by a set of metallic strips responsible for the 2DRR's electrical transduction, together with the grounded bottom metal plate. Similarly to the *overhang* electrodes used in Lamb wave devices [38], additional overhang corrugations (O.C.) in the piezoelectric layer (*e.g.* rods that are not covered by metal, thus not piezo-active) can also be used along the 2DRR's sides to avoid modal distortions due to the lateral stress-free boundaries of the 2DRR's plate. This ensures that all the unit-cells, even the ones not confining on both their sides with other unit-cells, exhibit the same transduction efficiency, allowing to maximize the achievable $k_t^2$. As we analytically demonstrated in [12], the corrugated 2DRR's structure enables unique acoustic dispersion features that do not exist in un-corrugated plates. For instance, it permits to generate stop-bands, inhibiting the acoustic propagation of real energy out of each 2DRR's unit-cell. This makes each unit-cell able to efficiently transduce its own *rod*-mode, with almost the same transduction efficiency as if it was not connected to any other unit-cells. More specifically, the corrugation in the piezoelectric film generates a reactive coupling between adjacent unit-cells that is strong enough to ensure a single frequency operation, but weak enough not to affect significantly the modal transduction within each unit-cell. Even more, the corrugated 2DRR's structure produces artificial and lithographically defined boundary conditions, allowing to squeeze the piezo generated displacement in the rods and near the interfaces between the piezo layer and the top metallic strips. Such a unique modal characteristic, together with the fact that most of the transduced acoustic energy can then be stored in resonator volumes bounded by lateral stress-free surfaces, makes the rods more compliant to both vertical and horizontal deformations. This provides the means to achieve a higher motional capacitance ($C_m$) than attained by conventional μARs relying on one or two-dimensional modes of vibration in un-corrugated plates. Regarding the 2DRR's excitation strategy, each *rod*-mode is transduced by generating a vertical electric field between the top and bottom metallic layers sandwiching each rod, and by leveraging both the $d_{31}$ and $d_{33}$ piezoelectric coefficients of the adopted piezoelectric layer. The extent to which the $d_{31}$ is used, ensuring a significant lithographic frequency tunability and a boost in the achievable $k_t^2$ due to a stronger lateral vibration in the rods, depends on the rods' geometry and material stack. It is also worth emphasizing that the use of a corrugated piezoelectric layer permits to conveniently reshape the electric field distribution within the entire 2DRR's volume so that just a negligible electric field exists within the trenches. This permits to avoid the undesired excitation of shear-modes from in-plane electric field components, even preventing $k_t^2$ reductions due to electrical energy being stored out of the rods (*i.e.* out of the 2DRR's active region).

Table I: Comparison among reported µARs with $f_{res}$ > 5GHz.

| Ref. | Material | CMOS Compatible | Freq. [GHz] | $k_t^2$ |
|---|---|---|---|---|
| [29] | LiNbO$_3$ | No | 5 | 26% |
| [30] | LiNbO$_3$ | No | 9.5 | 11% |
| [31] | LiNbO$_3$ | No | 5 | 9% |
| [32] | LiNbO$_3$ | No | 5 | 28% |
| [33] | AlN | Yes | 24 | 6% |
| [34] | AlN | Yes | 8.8 | 0.3% |
| [35] | AlN | Yes | 11 | 1.3% |
| [17] | AlScN | Yes | 9 | 5.3% |
| [36] | AlScN | Yes | 11.1 | 1.8% |
| [36] | AlScN | Yes | 19 | 3.5% |
| [37] | AlScN | Yes | 8.5 | 0.3% |
| **This work** | **AlScN** | **Yes** | **5.3** | **23.9%** |

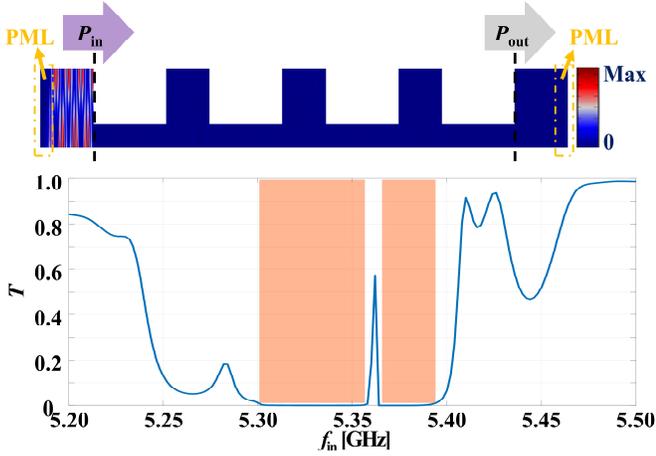

Figure 3: FEM simulated trend of $T$ vs. $f_{in}$ relative to the 2DRR's corrugated structure designed and built in this work. The identified stop-bands are highlighted in orange. The FEM simulated pressure distribution across the 2DRR's corrugated structure is also reported for a $f_{in}$ value (5.31 GHz) matching the $f_{res}$ value of the 2DRR built in this work.

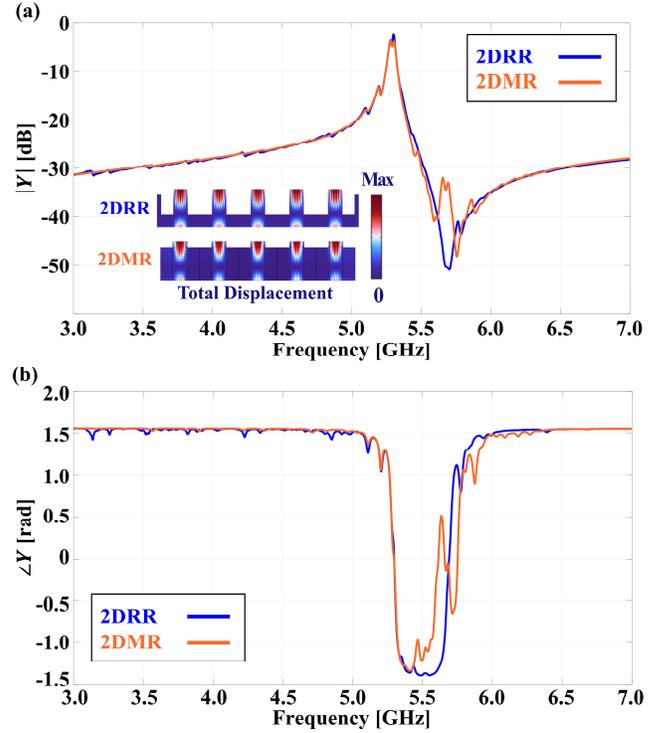

Figure 4: FEM simulated admittance responses [(a) magnitude ($|Y|$) and (b) phase ($\angle Y$)] relative to the 2DRR (blue-line) presented in this work and to a 2DMR (orange-curve) with the same geometry of the 2DRR but with no corrugation of the piezoelectric layer. The same $Q$ (200) was assumed in both the 2DRR and 2DMR simulations. The at-resonance modeshapes relative to the total displacement for both simulated devices are shown in the inset of (a).

### A. 2DRR –Design Flow

The 2DRR reported in this work was designed by relying on a purely acoustic finite-element simulation aiming at identifying the optimal cross-sectional dimensions leading to the generation of a stop-band around the desired operational frequency ($f_{tar}$~5.3 GHz). In order to do so and in line with what previously found [12][14], we assumed any rod-modes to have a resonance frequency matching closely the $f_{res}$ value of the thickness-extensional (TE) mode relative to the rods' material stack. By doing so and after selecting a specific material composition for the active region of the targeted 2DRR prototype, we were able to analytically find the $T_r$ value and the thicknesses of the two 2DRR's metal layers ($T_{bot}$=20/50 nm and $T_{top}$=100 nm) resulting into an $f_{res}$ value equal to $f_{tar}$. Later, we identified an optimal combination of rod-width ($W_r$), unit-cell width ($W_u$) and $T_l$ values, allowing to generate a stop-band around $f_{tar}$. In order to do so, we built a 2D simulation framework in COMSOL to investigate the acoustic transmission properties of the 2DRR's corrugated structure. This relied on an input pressure source (with frequency $f_{in}$), generating an acoustic power ($P_{in}$) longitudinally flowing through the left side of a 2DRR structure formed by 5 unit-cells (the same number used for the 2DRR built

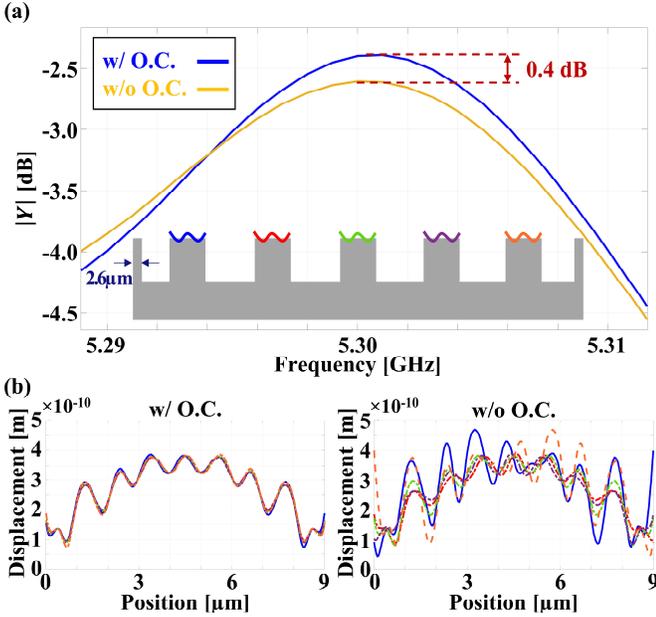

Figure 5: (a) At-resonance FEM simulated admittance responses ($Y$) (blue-line) of the 2DRR presented in this work (*e.g.* with the overhang corrugations, O.C.) and of the corresponding design without O.C. (yellow-line). The same $Q$ (200) has been considered in both simulations. Evidently, the 2DRR reported in this work shows a *0.4 dB* boost in $|Y(f_{res})|$ due to a *5%* larger $k_t^2$. The total displacement magnitude distribution across each rod's top surface is also reported for both devices (b), demonstrating how the O.C permit to suppress the modal distortion originated from the lateral stress-free boundaries

in this work). A power probe was used to quantify the acoustic power level ($P_{out}$) reaching the right side of the 2DRR structure, allowing to compute a *transmission coefficient* ($T$, equal to $|P_{out}/P_{in}|$) capturing the ability to transmit real power across the 2DRR's corrugated structure. Perfectly-Matched-Layers (PMLs) were also employed in our simulations to prevent undesired reflections due to acoustic scattering occurring at the outer edges of a bounded simulated geometry, which would otherwise degrade the accuracy and reliability of our results. The FEM simulated trend of $T$ *vs.* $f_{in}$ is reported in Figure 3, together with a schematic view summarizing the main features of our simulation framework. As evident, the designed 2DRR structure exhibits a series of nearly contiguous stop-bands for $f_{in}$ values close to $f_{tar}$, demonstrating that the interaction of the *local* rod resonances with the acoustic propagation characteristics of the trenches can indeed inhibit the propagation of real energy across the 2DRR structure. In other words, within the stop-bands, all the wave-vectors relative to the acoustic propagation through the trenches are imaginary. As a result, for $f_{in}$ values included in any stop-bands, there is no real power internally produced within each rod flowing out of the corresponding unit-cell, ensuring that just a weak and reactive coupling exists between adjacent unit-cells. In order to provide a visual representation of the acoustic behavior of the 2DRR structure operating within any one of its stop-bands, we report in Figure 3 the simulated modeshape of the magnitude of the acoustic pressure ($|p|$) generated across the 2DRR structure by the employed pressure source for a $f_{in}$ (*5.31 GHz*) coinciding with the measured $f_{res}$ value of the reported 2DRR device. Evidently and as expected, $|p|$ decays exponentially with the distance from the pressure source, leading to an almost nulled $T$ value.

After relying on purely acoustic FEM simulations to design the corrugated structure, we also simulated through FEM the electromechanical response of the resulting 2DRR. While doing so, two overhang corrugations of the piezoelectric layers, identified through FEM so as not to require additional fabrication steps, were added along the 2DRR's lateral sides to avoid modal distortions originated from the 2DRR's outer stress-free boundaries, further enhancing the achievable $k_t^2$. In order to do so, we applied a continuous-wave (CW) input voltage between the 2DRR's top and the bottom metal layers, with frequency ranging between *3 GHz* and *7 GHz*. Then, arbitrarily assuming a mechanical quality factor ($Q_m$) of *200*, we extracted the 2DRR's simulated admittance ($Y$, see Figure 4), Next, we extracted the 2DRR's expected $k_t^2$ by fitting the simulated $Y$ trend through a Modified-Butterworth-Van Dyke (MBVD) model [39]. As evident, the results of our FEM simulations clearly show that the designed 2DRR can simultaneously achieve a high $k_t^2$ (*~20%*), as well as an almost spurious-free electrical response. It is also worth emphasizing that the 2DRR reported here relies on a $W_r$ value that is much larger than $T_{tot}$, differently from our recent AlN prototype [12] using a $W_r$ value close to the total thickness of the piezoelectric layer within the rods ($T_{tot}$, equal to $T_r+T_l$). This allows to maintain a relaxed lithographic resolution, yet limiting the maximum lithographic frequency tunability ($df_{res}/dW_r \sim 9$ MHz/μm) and permitting to just partially leverage the AlScN $d_{31}$ piezoelectric coefficient. Nevertheless, the 2DRR's corrugated structure allows to focus most of the piezoelectrically generated acoustic energy under each rod and near the interfaces between the piezoelectric layer and the top metal layer (see the modeshape in Figure 4), thus in those portions of the 2DRR's active region laterally bounded by closely spaced stress-free surfaces. This ensures that the highest possible transduction efficiency can be achieved, given the applied purely vertical electric field distribution. To further demonstrate the advantage in terms of $k_t^2$ introduced by the 2DRR's topology, we simulated the admittance of the Two-Dimensional-Mode-Resonator (2DMR [5]) attained from our designed 2DRR when no corrugation of the AlScN layer is considered and when assuming the same $Q$ used for the 2DRR simulation. As evident from its *1.4 dB* lower resonant admittance peak, the 2DMR device exhibits a *17%* lower $k_t^2$ than the designed 2DRR. This is mainly due to the fact that the top-metallized 2DMR's regions, each one responsible for the transduction of a distinct $S_l$ Lamb wave mode [40], are separated by resonator portions that operate well below the cut-off of the $S_l$ mode. As a result, the top-metallized 2DMR's regions vibrate as they were clamped along their lateral sides, contrary to what happens in 2DRRs where most of the acoustic energy is stored in device portions limited by stress-free boundaries. Even more and as expected, the simulated 2DMR's response is more impacted by close-to-resonance spurious-modes than the designed 2DRR. Finally, while we aim to provide a detailed analysis of the effect of the overhang corrugations in a future more dedicated publication, here we show how such device features impact the device performance by comparing simulated $k_t^2$ and modeshapes of 2DRRs, with and without them. As evident from Figure 5, the introduction of the overhang corrugations permits to suppress the modal distortions otherwise affecting the outer unit-cells of the designed 2DRR, preventing the achievement of their maximum transduction efficiency. In fact, as verified through FEM, the introduction of the overhang corrugations allows a ~5% $k_t^2$ improvement with respect to the conventional case where no overhang corrugations are employed, without complicating the fabrication flow.

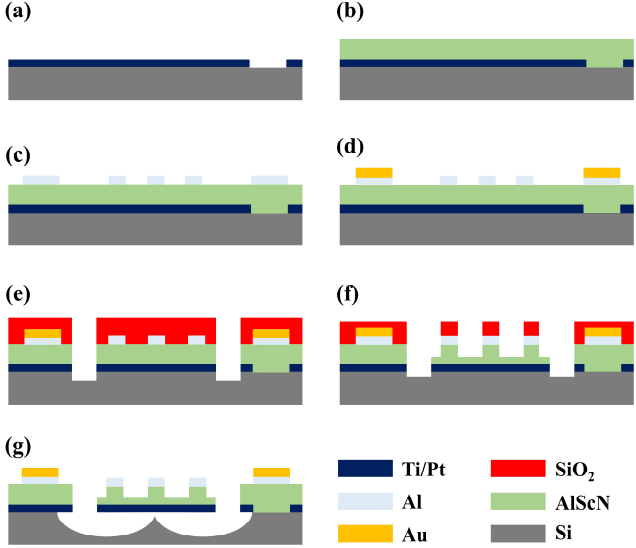

Figure 6: Schematic view of the fabrication flow used to build the reported 2DRR. (a) We started with the deposition of Ti/Pt. The portion under the probing pads was then removed through ICP RIE. (b) Then, we performed a reactive co-sputtering of Al and Sc in a $N_2$ atmosphere, to form the $Al_{0.76}Sc_{0.24}N$ film. (c) Next, we sputtered and patterned Al to form the metallic strips on top of the rods. (d) Similarly, a layer of Au was formed to cover the contact pads and routing. (e) Later, we created the release pits on the 2DRR's sides by dry etching, after depositing and patterning a PECVD-formed *2 µm*-thick $SiO_2$ layer used as hard mask. (f) Then, the same $SiO_2$ layer was patterned a second time for the AlScN partial-etch required to form the rods. (g) Finally, we removed the remaining $SiO_2$ by fluorine based ICP-RIE and we released the device through a $XeF_2$ isotropic etch.

### III. FABRICATION PROCESS

The AlScN 2DRR prototype reported in this work was fabricated on a low resistivity silicon wafer by using the process flow described in Figure 6. We first deposited and patterned a *20/50 nm*-thick Ti/Pt film. Then, we deposited a *500 nm*-thick $Al_{0.76}Sc_{0.24}$N-film by using a set of optimized deposition conditions that will be discussed in details in Section IV-a. Later, we deposited and patterned a *100 nm*-thick Al layer to form the metallic strips on top of each rod. Next, we deposited and patterned a *250 nm*-thick Au layer over the probing and routing areas to reduce the electrical loading. This step is critical, given the 2DRR's low impedance (*25 Ω*) and its expected motional resistance ($R_m$) lower than *1 Ω*. Then, the 2DRR's outer boundaries were formed by simultaneously etching both AlScN and Ti/Pt through an Inductively Coupled Plasma Reactively Ion Etching (ICP-RIE). This also allowed to generate the release pits for the structural release of the 2DRR device. Another ICP-RIE step was then run to form the trenches. Both etching steps relied on the same $SiO_2$ hard mask, patterned twice, and on an optimized recipe discussed in Section IV-b. Finally, the device was structurally released through a $XeF_2$ isotropic etch. It is worth emphasizing that the achievement of a grounded bottom metal plate in the 2DRR's active region has not been attained through the formation of vias. Instead, we engineered the routing and shape of the Ground-Signal-Ground pads to capacitively couple, at RF, the Ti/Pt plate at the bottom of the 2DRR to ground, simplifying the fabrication process and avoiding the introduction of additional ohmic losses due to vias.

### IV. ALSCN PROCESSING AND CHARACTERIZATION

Several challenges exist to deposit and etch AlScN films for micro- and nano- electromechanical (MEM/NEM) devices. While several groups have been recently looking at ways to epitaxially grow *nm*-thick AlScN films [41][42], sputtering remains the most adequate deposition technique when AlScN films with thicknesses in the hundreds of nanometers or more are needed. Therefore, in this work, we relied on sputtering to deposit the $Al_{0.76}Sc_{0.24}N$ layer used by the reported 2DRR prototype. Since the quality of sputtered and etched AlScN films is very sensitive to the adopted deposition and etching processes, proper procedures must be identified and followed to achieve patterned *c*-oriented AlScN films with the best possible crystalline orientation and with sharp sidewalls. Furthermore, the same procedures must give a low residual stress as well as a low percentage of *abnormally oriented grains* (AOGs). AOGs are irregularities in the crystalline structure of sputtered AlScN films, whose density tends to quickly grow with the adopted $Sc_\%$ value. Attaining a low AOG-density has shown to be fundamental to really leverage the superior electromechanical performance of highly-doped AlScN resonators [43]. Even more, the selection of higher $Sc_\%$ values makes it significantly more challenging to etch AlScN rather than AlN, especially if spatially consistent and sharp enough sidewall profiles are needed to prevent acoustic performance degradations due to mode conversion.

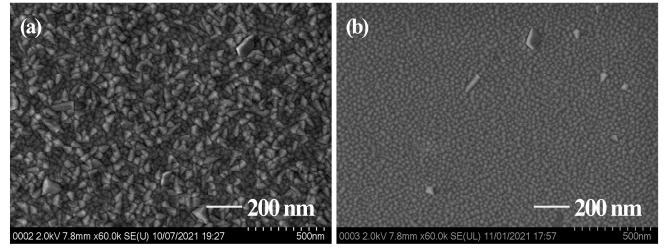

Figure 7: a-b) Scanned Electron Microscope (SEM) top-view pictures of two *500 nm*-thick $Al_{0.76}Sc_{0.24}N$ films deposited on top of two identical Ti/Pt stacks sputtered by using different temperatures during the deposition of the Ti layer. In particular: a) refers to a Ti layer deposited by using a chuck temperature of *50°C*; b) refers to a Ti layer deposited by using a chuck temperature of *500°C*.

Table II: Optimized Ti/Pt deposition recipe

| Step | Parameter | Value |
|---|---|---|
| Ti Deposition | Power to Ti Target | 1000 W |
| | Ar Gas flow | 45 sccm |
| | Target to Substrate Height | 30 mm |
| | Substrate Temperature | 500°C |
| | Base Pressure | < 9E-8 mbar |
| Pt Deposition | Power to Pt Target | 850 W |
| | Ar Gas flow | 40 sccm |
| | Target to Substrate Height | 20 mm |
| | Substrate Temperature | 500°C |
| | Base Pressure | < 9E-8 mbar |

*A. AlScN Deposition*

We deposited a *500 nm*-thick $Al_{0.76}Sc_{0.24}N$ film utilizing an EVATEC CLUSTERLINE® 200 MSQ multi-source system. In particular, we relied on a DC-pulsed co-sputtering process using Al and Sc 4-inch targets (see Figure 6-b). Several deposition parameters, such as chuck height, $N_2$ flow and chuck temperature impact the AlScN's crystalline quality. Also, the AOG density highly depends on the material stack on which the AlScN film is grown. For instance, a Ti/Pt stack was found, following the recipes in Table II, to lead to fewer AOGs during the growth of AlScN films than other metals or metal-stacks. Nevertheless, we found that the AlScN AOG density attained when using Ti/Pt heavily depends on the chuck temperature during the Ti deposition. In this regard, we show (Figure 7) two Scanned-Electron-Microscope (SEM) pictures of *500 nm*-thick $Al_{0.76}Sc_{0.24}N$ layers, deposited on top of *20 nm/50 nm* Ti/Pt stacks by using the same optimized deposition recipe (see Table III) but relying on different deposition conditions for the Ti layer. In particular, Figure 7-a shows a top-view of one of the sputtered $Al_{0.76}Sc_{0.24}N$ films when setting the chuck temperature during the Ti deposition at *50°C*. Evidently, a large density of AOGs was found in this case. A much more favorable AOG density was instead attained for the second $Al_{0.76}Sc_{0.24}N$ film, sputtered on top of an identical Ti/Pt stack, yet attained by setting the chuck temperature to *500°C* during the Ti deposition. As shown in Figure 7-b, the increase of the chuck temperature during the Ti deposition allowed us to largely reduce the AlScN AOG density. Such improvement is likely due to the fact that a Ti film deposited at higher temperatures can diffuse less into the Pt layer during the AlScN deposition. In any case, the minimization of the AlScN AOG density attained in this work made it possible to achieve superior quality in the AlScN film, allowing to really exploit the inherent high $k_t^2$ of AlScN 2DRRs. The optimized AlScN film was characterized in terms of both its stoichiometric composition and its rocking curve. The former was assessed by relying on Energy-Dispersive Spectroscopy (EDS). As shown in Figure 8-a, EDS provided us with the elementary composition of the deposited AlScN film on top of Ti/Pt and silicon. From the detected atomic percentages of Sc and Al, we were able to confirm a Sc concentration of *~24%*, matching closely our initially targeted value. The degree of crystallinity exhibited by our sputtered AlScN film was instead evaluated through an XRD measurement (see Figure 8-b), wherein the rocking curve extracted from the omega scan shows a Full-Width-at Half-Maximum (FWHM) of *1.9°*. Based on the optimized process flow we identified to deposit good-quality AlScN films not impacted by AOGs, it is important to emphasize that the 2DRRs' ability to achieve high $k_t^2$ values without patterning the bottom metal layer is fundamental, especially when targeting resonance frequencies in the SHF range. In fact, it prevents from having steps in the piezoelectric film locally perturbing the AlScN growth in the proximity of patterned bottom features and heavily degrading the achievable electromechanical performance, especially when operating at *>1 GHz* frequencies.

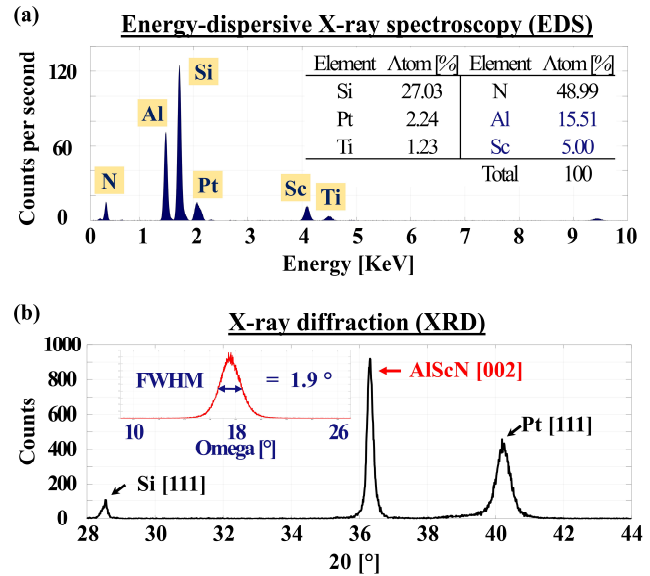

Figure 8: a) EDS result for the $Al_{0.76}Sc_{0.24}N$ film, deposited on top of the *20 nm/50 nm* Ti/Pt stack used for the 2DRR reported in this work. This shows an AlScN elementary composition of *~24%*, matching closely our targeted value; b) XRD result for the same $Al_{0.76}Sc_{0.24}N$ film. Clearly, the θ-2θ goniometer shows a peak at *36.2°*, indicating that the deposited AlScN film has a *c*-oriented crystalline structure. Also, as evident from the inset, the Full-Width-at Half-Maximum (FWHM) relative to the rocking curve of the same AlScN film is *1.9°*, which proves that an AlScN film with an excellent crystalline structure has been grown. It is worth mentioning that the two remaining XRD peaks, at *28.5°* and *40.1°*, are associated to the Si substrate and to the bottom Pt, respectively.

Table III: Optimized $Al_{0.76}Sc_{0.24}N$ deposition recipe

| Parameter | Value |
|---|---|
| Power to Al Target | 1000 W |
| Power to Sc Target | 450 W |
| $N_2$ Gas flow | 20 sccm |
| Target to Substrate Height | 33 mm |
| Substrate Temperature | 350 °C |
| Base Pressure | 9E-8 mbar |

Table IV: Optimized $Al_{0.76}Sc_{0.24}N$ etching recipe

| Parameter | Value |
|---|---|
| ICP Power | 600 W |
| RF Bias Power | 300 W |
| $Cl_2$ Gas flow | 10 sccm |
| $BCl_3$ Gas flow | 6 sccm |
| Ar Gas Flow | 28 sccm |
| Chamber Pressure | 10 mT |

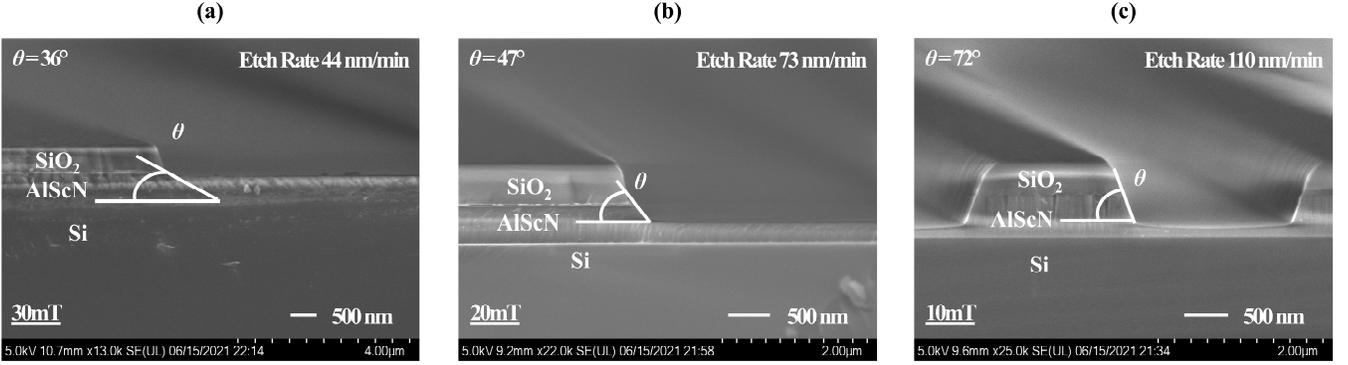

Figure 9: a-c) Scanned Electron Microscope (SEM) cross-sectional-view pictures of three *500 nm*-thick Al$_{0.76}$Sc$_{0.24}$N films (see Figure 8), deposited on top of a *20/50 nm* Ti/Pt stack, after being etched through an ICP RIE step wherein the chamber pressure was kept at *30 mT* (a), *20 mT* (b) and *10 mT* (c). For each one of these etching conditions, we report the corresponding achieved etching rate and AlScN sidewall angle.

Even more, having portions of the AlScN film grown on Si and others grown on Pt, within the same resonator's active region, would make it extremely more challenging to control the AOG density across the entire resonator's volume. In fact, substantially different recipes are needed to minimize the AOG density of AlScN films deposited on Si or Pt. Even more, due to the high chuck temperature required during the Ti/Pt deposition, a patterning of the bottom Ti/Pt layer cannot be made through lift-off but requires etching Ti/Pt through a physical-etching mechanism, resulting into even more challenges in the achievement of good quality AlScN films due to the increased roughness of Si such etch mechanism would certainly cause.

### B. AlScN Etching

As we mentioned in Section IV, there are two AlScN etching steps involved in the fabrication of the reported 2DRR. One step permits to create the release pits and to define the outer edges of the suspended piezoelectric membrane after its structural release. The other step consists, instead, in a partial etching of the AlScN film, aiming at generating the 2DRR's corrugated structure. For both steps, the most important goal is the generation of sharp etching sidewalls, preventing any performance degradations due to acoustic mode conversion. As discussed in Section IV, both etching steps were run by using ICP-RIE, a technique offering a combination of both physical and chemical etching that is particularly effective whenever thin-films with well-aligned crystalline structures need to be etched. Therefore, we benefitted from the good crystalline orientation attained by our deposited AlScN film (see Figure 8) even during the characterization of the AlScN etching recipe. We started by setting the Cl$_2$/BCl$_3$/Ar gas flow composition to *10/6/28 sccm*, as well as the ICP/RF-Bias to *600/300 W*. This set-up has recently showed to be effective in etching highly doped AlScN films. Soon, we realized that the chamber pressure (*P*) during the etching process was a critical parameter to control the AlScN sidewall angle ($\theta$). In particular, we verified that the lateral etching responsible for the AlScN sidewall profile can be largely reduced by lowering *P* during the etch, even allowing to exploit higher etching rates. In Figure 9, we report three Scanned Electron Microscope (SEM) pictures showing the sidewall profile of three pieces of the Al$_{0.76}$Sc$_{0.24}$N film discussed in Figure 8 when using the etching parameters reported in Table IV, as well as three different *P* values (*30 mT*, *20 mT* and *10 mT*). Evidently, relying on the lowest *P* value permits to achieve the best $\theta$ (*72°*), as well as the highest etching rate (*110 nm/min*, thus *2.5* times faster than the one attained when relying on a *P* value of *30 mT*). Even more, the amount of physical etching ensured by the adoption of such a low *P* value suffices to etch even the Ti/Pt layer under the AlScN film. The optimized parameters of the etching recipe we developed in this work are summarized in Table IV.

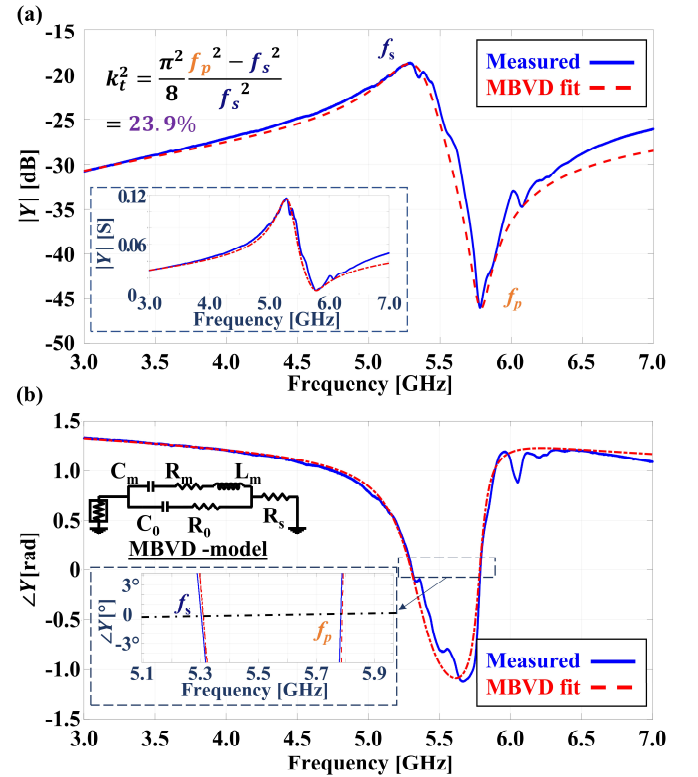

Figure 10: Experimentally extracted *Y* for the reported 2DRR [(a) magnitude (|*Y*|) and (b) phase (∠*Y*)]. The blue curves represent the measured admittance response while the dashed lines show the corresponding MBVD fitted response, which has been used to extract all the electromechanical parameters listed in Table V. A linear plot of |*Y*| (in Siemens) is also shown in the inset of (a). A zoomed version of ∠*Y* around the two resonances is shown in the inset of (b).

## V. Experimental Results

After building the reported 2DRR through the fabrication process discussed in Section III, we characterized its electromechanical performance by means of a conventional RF characterization. Such characterization targeted the extraction of the measured 2DRR's admittance *vs.* frequency (Figure 10) through a Vector Network Analyzer (VNA). The measured 2DRR's response was then fitted by using an MBVD model to extract its electromechanical parameters, including: the statistic capacitance ($C_0$), the $k_t^2$ value, $Q_m$, $R_m$, the series resistance due to electrical routing ($R_s$), the parallel resistance ($R_0$) due to dielectric losses in the substrate, the mechanical figure of merit ($FoM^{(m)} = k_t^2 \cdot Q_m$) and the unloaded figure-of-merit ($FoM$). $FoM$ was extracted from the measured admittance at the 2DRR's parallel resonance frequency in favor of a more general performance evaluation, independent of the $C_0$ value selected for the reported 2DRR prototype. All the fitted parameters are listed in Table V. As evident, the built 2DRR operates at *5.31 GHz* and shows a record-high $k_t^2$ of *23.9%*, while being characterized by a capacitive impedance with magnitude approximately equal to *24 Ω*. It is worth emphasizing that such a high $k_t^2$ value exceeds what expected from our FEM simulation (~*20%*). Thanks to its superior $k_t^2$ and despite its relatively low $Q_m$ (*101*), the reported 2DRR shows a $FoM$ value of *12* and a $FoM^{(m)}$ value of *24*, one of the highest ones ever demonstrated for SHF AlScN microacoustic resonators. Furthermore, we found the resonance-to-antiresonance ratio of admittance to be *23.3*, which corroborates the extracted $FoM^{(m)}$. While we believe that a much higher $Q_m$ will be possible in the future through further design optimizations, the achievement of such an unprecedented $k_t^2$ makes it already possible to envision future high-order 2DRR's based acoustic filters with a sub-3dB insertion-loss (I.L) value and exhibiting an unprecedented ultra-wide BW (~*12%*, see Figure 11-(a,b)). This has been further demonstrated here by designing a *50 Ω*-matched 5[th]-order *5.3 GHz* 2DRR-based acoustic ladder filter, formed by resonators with the same electromechanical performance measured for the 2DRR reported in this work and by relying on a capacitance ratio ($r= C_0^P/C_0^S$, where $C_0^S$ and $C_0^P$ are the designed static capacitance of the series and shunt resonators, respectively) set to *3* (see Figure 11-(c)). As evident from Figure 11-(d), the simulated transmission ($S_{21}$) of the designed ladder filter clearly shows that the 2DRR's electromechanical performance demonstrated in

Table V: MBVD-Fitted electromechanical parameters for the 2DRR reported in this work (see Figure 10)

| Parameter | $C_0$ [fF] | $R_m$ [Ω] | $R_s$ [Ω] | $R_0$ [Ω] |
|---|---|---|---|---|
| Value | 1250 | 0.7 | 7.7 | 1.5 |
| Parameter | $k_t^2$ | $Q_m$ | $FoM^{(m)}$ | $FoM$* |
| Value | 23.9 | 101 | 24 | 12 |

*Calculated as $FoM = (2\pi f_P \cdot C_0)/|Y(f_P)|$, where $f_P$ is the anti-resonance frequency and $|Y(f_P)|$ is the admittance magnitude at $f_P$ this work simultaneously enable an I.L of ~*2.5 dB*, a BW of *11.1%* and an out-of-band rejection exceeding *30 dB*.

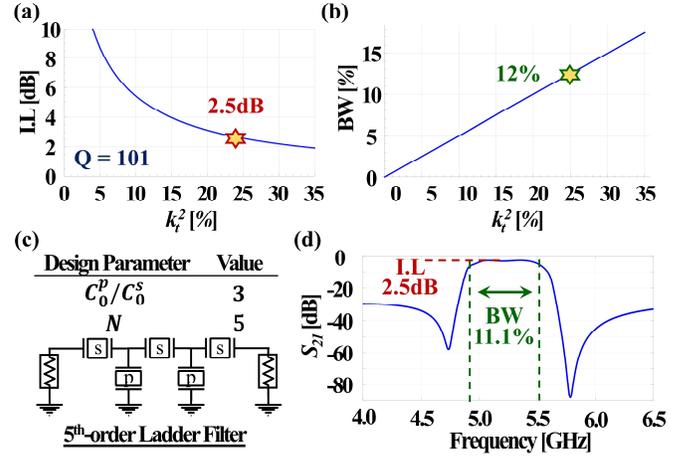

Figure 11: a-b) Simulated I.L (a) and BW (b) relative to a *50 Ω*-matched 5[th]-order *5.3 GHz* Al$_{0.76}$Sc$_{0.24}$N ladder filter, relying on a capacitance ratio equal to *3* and formed by properly sized 2DRRs with a quality factor equal to the $Q_m$ value demonstrated for the 2DRR reported in this work (101, see Table V) and with a $k_t^2$ varying between *0* and *35%*; c) Schematic view of a designed 5[th]-order *5.3 GHz* Al$_{0.76}$Sc$_{0.24}$N 2DRR-based ladder filter, relying on a capacitance ratio equal to *3* and on resonators with the same $k_t^2$ demonstrated in this work; d) Simulated transmission ($S_{21}$) for the filter in (c), when assuming the same electromechanical performance demonstrated for the 2DRR reported in this work.

## VI. Conclusions

In this Article, we reported on design, fabrication and measured performance of an Al$_{0.76}$Sc$_{0.24}$N Two-Dimensional-Resonant-Rods (2DRR) resonator exhibiting the highest electromechanical coupling coefficient ever reported for AlN or AlScN microacoustic resonant devices. The 2DRR reported here exhibits a resonance frequency of *5.31 GHz* and shows a $k_t^2$ of *23.9%*, leading to a mechanical figure-of-merit equal to *24*. The ability to achieve such an extraordinary $k_t^2$ in a device manufacturable through CMOS-compatible fabrication processes opens unprecedented scenarios towards the development of next generation low loss ultra-wideband (UWB) SHF acoustic filters for future *5G* communication systems.

## VII. Acknowledgment

This work was supported by the National Science Foundation (NSF) through the CAREER award (2034948). The authors wish to thank the staff of the George J. Kostas Nanoscale Technology and Manufacturing Research Center at Northeastern University and the staff of the Center for Nanoscale Systems at Harvard University for assistance in the device fabrication.

## VIII. References

[1] H. P. Loebl et al., "Solidly mounted bulk acoustic wave filters for the GHz frequency range," in *2002 IEEE Ultrasonics Symposium, 2002. Proceedings.*, Oct. 2002, vol. 1, pp. 919–923 vol.1. doi: 10.1109/ULTSYM.2002.1193546.

[2] R. C. Ruby, P. Bradley, Y. Oshmyansky, A. Chien, and J. D. Larson, "Thin film bulk wave acoustic resonators (FBAR) for wireless applications," in *2001 IEEE Ultrasonics Symposium. Proceedings. An International Symposium (Cat. No.01CH37263)*, Oct. 2001, vol. 1, pp. 813–821 vol.1. doi: 10.1109/ULTSYM.2001.991846.

[3] V. Yantchev and I. Katardjiev, "Thin film Lamb wave resonators in frequency control and sensing applications: a review," *J. Micromechanics Microengineering*, vol. 23, no. 4, p. 043001, Mar. 2013, doi: 10.1088/0960-1317/23/4/043001.

[4] C. Cassella, J. Segovia-Fernandez, and G. Piazza, "Segmented electrode excitation of aluminum nitride contour mode resonators to optimize the device figure of merit," in *2013 Transducers Eurosensors XXVII: The 17th International Conference on Solid-State Sensors, Actuators and Microsystems (TRANSDUCERS EUROSENSORS XXVII)*, Jun. 2013, pp. 506–509. doi: 10.1109/Transducers.2013.6626814.

[5] C. Cassella and G. Piazza, "AlN Two-Dimensional-Mode Resonators for Ultra-High Frequency Applications," *IEEE Electron Device Lett.*, vol. 36, no. 11, pp. 1192–1194, Nov. 2015, doi: 10.1109/LED.2015.2475172.

[6] Y. Zhu, N. Wang, C. Sun, S. Merugu, N. Singh, and Y. Gu, "A High Coupling Coefficient 2.3-GHz AlN Resonator for High Band LTE Filtering Application," *IEEE Electron Device Lett.*, vol. 37, no. 10, pp. 1344–1346, Oct. 2016, doi: 10.1109/LED.2016.2602852.

[7] C. Cassella, Y. Hui, Z. Qian, G. Hummel, and M. Rinaldi, "Aluminum Nitride Cross-Sectional Lamé Mode Resonators," *J. Microelectromechanical Syst.*, vol. 25, no. 2, pp. 275–285, Apr. 2016, doi: 10.1109/JMEMS.2015.2512379.

[8] C. Cassella and J. Segovia-Fernandez, "High $k_t^2$ Exceeding 6.4% Through Metal Frames in Aluminum Nitride 2-D Mode Resonators," *IEEE Trans. Ultrason. Ferroelectr. Freq. Control*, vol. 66, no. 5, pp. 958–964, May 2019, doi: 10.1109/TUFFC.2019.2903011.

[9] M. Rinaldi, C. Zuniga, C. Zuo, and G. Piazza, "Super-high-frequency two-port AlN contour-mode resonators for RF applications," *IEEE Trans. Ultrason. Ferroelectr. Freq. Control*, vol. 57, no. 1, pp. 38–45, Jan. 2010, doi: 10.1109/TUFFC.2010.1376.

[10] R. Ruby, P. Bradley, J. D. Larson, and Y. Oshmyansky, "PCS 1900 MHz duplexer using thin film bulk acoustic resonators (FBARs)," *Electron. Lett.*, vol. 35, no. 10, pp. 794–795, May 1999.

[11] C. Cassella, G. Chen, Z. Qian, G. Hummel, and M. Rinaldi, "Unprecedented Figure Of Merit In Excess Of 108 In 920 MHz Aluminum Nitride Cross-Sectional Lamé Mode Resonators Showing $k_t^2$ In Excess Of 6.2%," in *2016 Solid-State, Actuators, and Microsystems Workshop Technical Digest*, Hilton Head, South Carolina, USA, May 2016, pp. 94–97. doi: 10.31438/trf.hh2016.27.

[12] X. Zhao, L. Colombo, and C. Cassella, "Aluminum nitride two-dimensional-resonant-rods," *Appl. Phys. Lett.*, vol. 116, no. 14, p. 143504, Apr. 2020, doi: 10.1063/5.0005203.

[13] M. A. Caro et al., "Piezoelectric coefficients and spontaneous polarization of ScAlN," *J. Phys. Condens. Matter*, vol. 27, no. 24, p. 245901, Jun. 2015, doi: 10.1088/0953-8984/27/24/245901.

[14] J. E. A. Southin and R. W. Whatmore, "Finite element modelling of nanostructured piezoelectric resonators (NAPIERs)," *IEEE Trans. Ultrason. Ferroelectr. Freq. Control*, vol. 51, no. 6, pp. 654–662, Jun. 2004, doi: 10.1109/TUFFC.2004.1304263.

[15] Y. Zhu, N. Wang, G. Chua, C. Sun, N. Singh, and Y. Gu, "ScAlN-Based LCAT Mode Resonators Above 2 GHz With High FOM and Reduced Fabrication Complexity," *IEEE Electron Device Lett.*, vol. 38, no. 10, pp. 1481–1484, Oct. 2017, doi: 10.1109/LED.2017.2747089.

[16] N. Wang et al., "Over 10% of $k^2_{eff}$ Demonstrated by 2-GHz Spurious Mode-Free Sc0.12Al0.88N Laterally Coupled Alternating Thickness Mode Resonators," *IEEE Electron Device Lett.*, vol. 40, no. 6, pp. 957–960, Jun. 2019, doi: 10.1109/LED.2019.2910836.

[17] M. Park, Z. Hao, D. G. Kim, A. Clark, R. Dargis, and A. Ansari, "A 10 GHz Single-Crystalline Scandium-Doped Aluminum Nitride Lamb-Wave Resonator," in *2019 20th International Conference on Solid-State Sensors, Actuators and Microsystems Eurosensors XXXIII (TRANSDUCERS EUROSENSORS XXXIII)*, Jun. 2019, pp. 450–453. doi: 10.1109/TRANSDUCERS.2019.8808374.

[18] M. Moreira, J. Bjurström, I. Katardjev, and V. Yantchev, "Aluminum scandium nitride thin-film bulk acoustic resonators for wide band applications," *Vacuum*, vol. 86, no. 1, pp. 23–26, Jul. 2011, doi: 10.1016/j.vacuum.2011.03.026.

[19] A. Lozzi, E. Ting-Ta Yen, P. Muralt, and L. G. Villanueva, "Al0.83Sc0.17N Contour-Mode Resonators With Electromechanical Coupling in Excess of 4.5%," *IEEE Trans. Ultrason. Ferroelectr. Freq. Control*, vol. 66, no. 1, pp. 146–153, Jan. 2019, doi: 10.1109/TUFFC.2018.2882073.

[20] L. Colombo, A. Kochhar, C. Xu, G. Piazza, S. Mishin, and Y. Oshmyansky, "Investigation of 20% scandium-doped aluminum nitride films for MEMS laterally vibrating resonators," in *2017 IEEE International Ultrasonics Symposium (IUS)*, Sep. 2017, pp. 1–4. doi: 10.1109/ULTSYM.2017.8092076.

[21] Z. A. Schaffer, G. Piazza, S. Mishin, and Y. Oshmyansky, "Super High Frequency Simple Process Flow Cross-Sectional Lamé Mode Resonators in 20% Scandium-Doped Aluminum Nitride," in *2020 IEEE 33rd International Conference on Micro Electro Mechanical Systems (MEMS)*, Jan. 2020, pp. 1281–1284. doi: 10.1109/MEMS46641.2020.9056279.

[22] S. Rassay, D. Mo, C. Li, N. Choudhary, C. Forgey, and R. Tabrizian, "Intrinsically Switchable Ferroelectric Scandium Aluminum Nitride Lamb-Mode Resonators," *IEEE Electron Device Lett.*, vol. 42, no. 7, pp. 1065–1068, Jul. 2021, doi: 10.1109/LED.2021.3078444.

[23] M. Park and A. Ansari, "Epitaxial Al0.77Sc0.23N SAW and Lamb Wave Resonators," in *2020 Joint Conference of the IEEE International Frequency Control Symposium and International Symposium on Applications of Ferroelectrics (IFCS-ISAF)*, Jul. 2020, pp. 1–3. doi: 10.1109/IFCS-ISAF41089.2020.9234850.

[24] M. Schneider, M. DeMiguel-Ramos, A. J. Flewitt, E. Iborra, and U. Schmid, "Scandium Aluminium Nitride-Based Film Bulk Acoustic Resonators," *Proceedings*, vol. 1, no. 4, Art. no. 4, 2017, doi: 10.3390/proceedings1040305.

[25] G. Esteves et al., "Al0.68Sc0.32N Lamb wave resonators with electromechanical coupling coefficients near 10.28%," *Appl. Phys. Lett.*, vol. 118, no. 17, p. 171902, Apr. 2021, doi: 10.1063/5.0047647.

[26] C. Moe et al., "Highly Doped AlScN 3.5 GHz XBAW Resonators with 16% $k^2_{eff}$ for 5G RF Filter Applications," in *2020 IEEE International Ultrasonics Symposium (IUS)*, Sep. 2020, pp. 1–4. doi: 10.1109/IUS46767.2020.9251412.

[27] J. Wang, M. Park, S. Mertin, T. Pensala, F. Ayazi, and A. Ansari, "A Film Bulk Acoustic Resonator Based on Ferroelectric Aluminum Scandium Nitride Films," *J. Microelectromechanical Syst.*, vol. 29, no. 5, pp. 741–747, Oct. 2020, doi: 10.1109/JMEMS.2020.3014584.

[28] K. Umeda, H. Kawai, A. Honda, M. Akiyama, T. Kato, and T. Fukura, "Piezoelectric properties of ScAlN thin films for piezo-MEMS devices," in *2013 IEEE 26th International Conference on Micro Electro Mechanical Systems (MEMS)*, Jan. 2013, pp. 733–736. doi: 10.1109/MEMSYS.2013.6474347.

[29] Y. Yang, A. Gao, R. Lu, and S. Gong, "5 GHz lithium niobate MEMS resonators with high FoM of 153," in *2017 IEEE 30th International Conference on Micro Electro Mechanical Systems (MEMS)*, Jan. 2017, pp. 942–945. doi: 10.1109/MEMSYS.2017.7863565.

[30] X. Chen, M. A. Mohammad, J. Conway, B. Liu, Y. Yang, and T.-L. Ren, "High performance lithium niobate surface acoustic wave transducers in the 4–12 GHz super high frequency range," *J. Vac. Sci. Technol. B*, vol. 33, no. 6, p. 06F401, Nov. 2015, doi: 10.1116/1.4935561.

[31] T. Kimura et al., "A High Velocity and Wideband SAW on a Thin LiNbO3 Plate Bonded on a Si Substrate in the SHF Range," in *2019 IEEE International Ultrasonics Symposium (IUS)*, Oct. 2019, pp. 1239–1248. doi: 10.1109/ULTSYM.2019.8926065.

[32] S. Yandrapalli, S. E. K. Eroglu, V. Plessky, H. B. Atakan, and L. G. Villanueva, "Study of Thin Film LiNbO3 Laterally Excited Bulk Acoustic Resonators," J. Microelectromechanical Syst., vol. 31, no. 2, pp. 217–225, Apr. 2022, doi: 10.1109/JMEMS.2022.3143354.

[33] M. Hara et al., "Super-High-Frequency Band Filters Configured with Air-Gap-Type Thin-Film Bulk Acoustic Resonators," *Jpn. J. Appl. Phys.*, vol. 49, no. 7S, p. 07HD13, Jul. 2010, doi: 10.1143/JJAP.49.07HD13.

[34] G. Chen and M. Rinaldi, "High-Q X Band Aluminum Nitride Combined Overtone Resonators," in *2019 Joint Conference of the IEEE*

*International Frequency Control Symposium and European Frequency and Time Forum (EFTF/IFC)*, Apr. 2019, pp. 1–3. doi: 10.1109/FCS.2019.8856047.

[35] M. Assylbekova, G. Chen, G. Michetti, M. Pirro, L. Colombo, and M. Rinaldi, "11 GHz Lateral-Field-Excited Aluminum Nitride Cross-Sectional Lamé Mode Resonator," in *2020 Joint Conference of the IEEE International Frequency Control Symposium and International Symposium on Applications of Ferroelectrics (IFCS-ISAF)*, Jul. 2020, pp. 1–4. doi: 10.1109/IFCS-ISAF41089.2020.9234874.

[36] M. Park, J. Wang, R. Dargis, A. Clark, and A. Ansari, "Super High-Frequency Scandium Aluminum Nitride Crystalline Film Bulk Acoustic Resonators," in *2019 IEEE International Ultrasonics Symposium (IUS)*, Oct. 2019, pp. 1689–1692. doi: 10.1109/ULTSYM.2019.8925598.

[37] M. Rinaldi, C. Zuniga, and G. Piazza, "5-10 GHz AlN Contour-Mode Nanoelectromechanical Resonators," in *2009 IEEE 22nd International Conference on Micro Electro Mechanical Systems*, Jan. 2009, pp. 916–919. doi: 10.1109/MEMSYS.2009.4805533.

[38] S. Gong and G. Piazza, "Figure-of-Merit Enhancement for Laterally Vibrating Lithium Niobate MEMS Resonators," in IEEE Transactions on Electron Devices, vol. 60, no. 11, pp. 3888-3894, Nov. 2013, doi: 10.1109/TED.2013.2281734.

[39] J. D. Larson, P. D. Bradley, S. Wartenberg, and R. C. Ruby, "Modified Butterworth-Van Dyke circuit for FBAR resonators and automated measurement system," in *2000 IEEE Ultrasonics Symposium. Proceedings. An International Symposium (Cat. No.00CH37121)*, Oct. 2000, vol. 1, pp. 863–868 vol.1. doi: 10.1109/ULTSYM.2000.922679.

[40] C. Cassella and M. Rinaldi, "On the Origin of High Couplings Two-Dimensional Modes of Vibration in Aluminum Nitride Plates," in 2018 IEEE International Frequency Control Symposium (IFCS), May 2018, pp. 1–3. doi: 10.1109/FCS.2018.8597568.

[41] J. Wang, M. Park, and A. Ansari, "High-Temperature Acoustic and Electric Characterization of Ferroelectric Al0.7Sc0.3N Films," *J. Microelectromechanical Syst.*, pp. 1–7, 2022, doi: 10.1109/JMEMS.2022.3147492.

[42] G. Schönweger *et al.*, "From Fully Strained to Relaxed: Epitaxial Ferroelectric Al1-xScxN for III-N Technology," *Adv. Funct. Mater.*, vol. n/a, no. n/a, p. 2109632, doi: 10.1002/adfm.202109632.

[43] C. S. Sandu *et al.*, "Abnormal Grain Growth in AlScN Thin Films Induced by Complexion Formation at Crystallite Interfaces," *Phys. Status Solidi A*, vol. 216, no. 2, p. 1800569, 2019, doi: 10.1002/pssa.201800569.

[44] X. Zhao and C. Cassella, "A Comparative Study on the Performance of Aluminum Nitride Thickness and quasi-Thickness Extensional Mode Resonators," in 2020 IEEE International Ultrasonics Symposium (IUS), Sep. 2020, pp. 1–4. doi: 10.1109/IUS46767.2020.9251735.